\documentclass[12pt,preprint]{aastex}
\usepackage{natbib}
\usepackage{graphicx,rotating}
\usepackage{amssymb}
\usepackage{epstopdf}
\usepackage{amsmath}
\usepackage{natbib}
\usepackage{lscape}
\usepackage{adjustbox,lipsum}
\newcommand{\beq}{\begin{equation}}
\newcommand{\eeq}{\end{equation}}
\def\etal{{\sl et~al.~}}
\newcommand{\HDs}{{HD\,202206~}}
\newcommand{\HD}{{HD\,202206}}

\newcommand{\HST}{{\it HST}}
\newcommand{\HSTs}{{\it HST~}}
\newcommand{\HIP}{{\it Hipparcos}}
\newcommand{\G}{{\it Gaia}}
\newcommand{\kms}{km s$^{-1}$~}

\newcommand{\msini}{$\cal{M}$\,sin\,{i}~}

\newcommand{\m}{$\cal{M}$}
\newcommand{\mjup}{$\cal{M}_{\rm Jup}~$}
\newcommand{\mjupe}{$\cal{M}_{\rm Jup}$}
\newcommand{\msun}{$\cal{M}_{\odot}~$}
\newcommand{\msune}{$\cal{M}_{\odot}$}

\received{}
\revised{}
\accepted{}

\shorttitle{The HD202206 System}
\shortauthors{Benedict \& Harrison}

\begin{document}
\bibliographystyle{/Active/my2}

\title{HD 202206 : A Circumbinary Brown Dwarf System\footnote{Based on observations made with the NASA/ESA Hubble Space Telescope, obtained at the Space Telescope Science Institute, which is operated by the Association of Universities for Research in Astronomy, Inc., under NASA contract NAS5-26555. } }

\author{ G.\ Fritz Benedict\altaffilmark{2} and Thomas E. Harrison\altaffilmark{3}}

\altaffiltext{2}{McDonald Observatory, University of Texas, Austin, TX 78712}

\altaffiltext{3}{Department of Astronomy, New Mexico State University, Box 30001, MSC 4500, Las Cruces, NM 88003-8001}




\begin{abstract}
With {\it Hubble Space Telescope} Fine Guidance Sensor astrometry and previously published radial velocity measures we explore the exoplanetary system \HD . Our modeling results in a parallax, $\pi_{abs} = 21.96\pm0.12$ milliseconds of arc, a mass  for \HDs B of \m$_B = 0.089^{ +0.007}_{-0.006}$ \msune, and a mass  for \HDs c of \m$_c = 17.9 ^{ +2.9}_{-1.8}$ \mjupe. \HD~is a nearly face-on G+M binary orbited by a brown dwarf. The system architecture we determine supports past assertions that stability requires a 5:1 mean motion resonance (we find a period ratio, $P_c/P_B = 4.92\pm0.04$) and coplanarity (we find a mutual inclination, $\Phi = 6\arcdeg \pm 2\arcdeg$).
\end{abstract}




%

\section{Introduction}

 We present our astrometric investigation of \HD, yielding parallax, proper motion, and measures of the perturbations due to companions \HDs B and c. Companion masses and the \HDs system architecture are the ultimate goals.
\cite{Udr02}  first reported on the discovery of a possible exoplanetary companions to \HD, using Doppler spectroscopy. \cite{Cor05} found a second companion with additional radial velocity (RV) data.  The title of the \cite{Cor05} paper, ``A pair of planets around HD 202206 or a circumbinary planet?", indicated a need for astrometry capable of measuring inclination.

{The issue of the stability of the \HDs system has engaged dynamicists from its original discovery as multi-component \citep{Cor05}. Using a symplectic integrator \citep{Las01} and frequency analysis \citep{Las90}, Correia et al. concluded that islands of stability (in longitude of periastron - semi-major axis space) existed for a system in 5:1 mean motion resonance (MMR). \cite{Cou10} incorporated the \cite{Cor05}  RV and additional RV data into their analysis of the stability of the \HDs system. Using similar tools they found that a 5:1 MMR was most likely to provide stability, and also found increased stability for coplanar system architecture. According to a stability criterion devised by \cite{Pet15},
the \HDs system is unstable, unless coplanar and in a MMR. Critically missing in all of these dynamical analyses are the true masses of each component.
}

 With only RV the inferred masses depend on their orbital inclination angle, $i$, providing minimum mass values $\cal{M}$$_b\,sin\,i$=17.4\mjup and $\cal{M}$$_c\,sin\,i$=2.44\mjupe. Hence, we included this system in an \HST~proposal \citep{Ben07c} to carry out astrometry using the Fine Guidance Sensors (FGS). They produced astrometry with which to establish the architectures of several promising candidate systems, all relatively nearby with companion \msini values and periods suggesting measurable astrometric amplitudes. Table~\ref{tbl-STAR} contains previously determined information and sources for the host star subject of this paper, \HD. 

In this paper we follow analysis procedures previously employed for the putative (now established) exoplanetary systems $\upsilon$ And \citep{McA10}, HD 136118 \citep{Mar10}, HD 38529 \citep{Ben10}, and HD 128311 \citep{McA14}. As summarized in \cite{Ben17},  perturbation amplitudes measured with the FGS have rarely exceeded a few milliseconds of arc (hereafter, mas).

Section~\ref{Ap1} describes our modeling approach, combining FGS astrometry with previously available ground-based RV.  We present the results of this modeling, component masses and mutual inclination  in Section~\ref{MMI}, and briefly discuss these results (Section~\ref{Disc}) in the context of dynamical explorations of the overall stability of the HD 202206 system. Lastly, in Section~\ref{Summ} we summarize our findings.

\section{Parallax, Proper Motion, and Companion Masses for\\ HD 202206}  \label{Ap1}

For this study astrometric measurements came from Fine Guidance Sensor\,1r (FGS\,1r), an upgraded FGS installed in 1997 during the second \HSTs servicing mission. 
It provided superior fringes from which to obtain target and reference star 
positions  \citep{McA02}. 

We utilized only the fringe tracking mode (POS-mode; see Benedict et al. 2017 for a review of this technique, and Nelan et al. 2015 for further details)\nocite{Nel15a,Ben17} in this investigation. \ {POS mode observations of a star 
have a typical duration of 60 seconds, during which over two thousand individual position measures are collected. The astrometric centroid is estimated by choosing the median measure, after filtering large outliers (caused by cosmic ray hits and particles trapped by the Earth's magnetic field). The standard deviation of the measures provides a measurement error. We refer to the aggregate of astrometric centroids of each star secured during one visibility period as an ``orbit". }
Because one of the pillars of the scientific method involves reproducibility, we present a complete ensemble of time-tagged  \HDs and reference star astrometric measurements, OFAD\footnote{The Optical Field Angle Distortion (OFAD) calibration \ {\citep{McA06}} reduces \HST~and FGS as-built optical distortions of order 2 seconds of arc to less than one mas in the center of the FGS field of regard. \ {This level of correction persists for average radial distances from FGS FOV center $<r>\leq100"$, and is a reason the parallax error for $\kappa$ Pav ($\pm$0.28 mas, $<r>=117"$) is over twice that of RR Lyr ($\pm$0.13 mas, $<r>=44"$)\citep{Ben11}.}}- and intra-orbit drift-corrected, in Table~\ref{tbl-DATA}, along with calculated parallax factors in Right Ascension and Declination. These data, collected from 2007.5 to 2010.4,  in addition to providing material for confirmation of our results, might ultimately be combined with $Gaia$ measures, significantly extending the time baseline of astrometry, thereby improving proper motion and perturbation characterization.

\subsection{\HDs Astrometric Reference Frame}  \label{AstRefs}
\begin{center}
\begin{figure}
\includegraphics[width=6in]{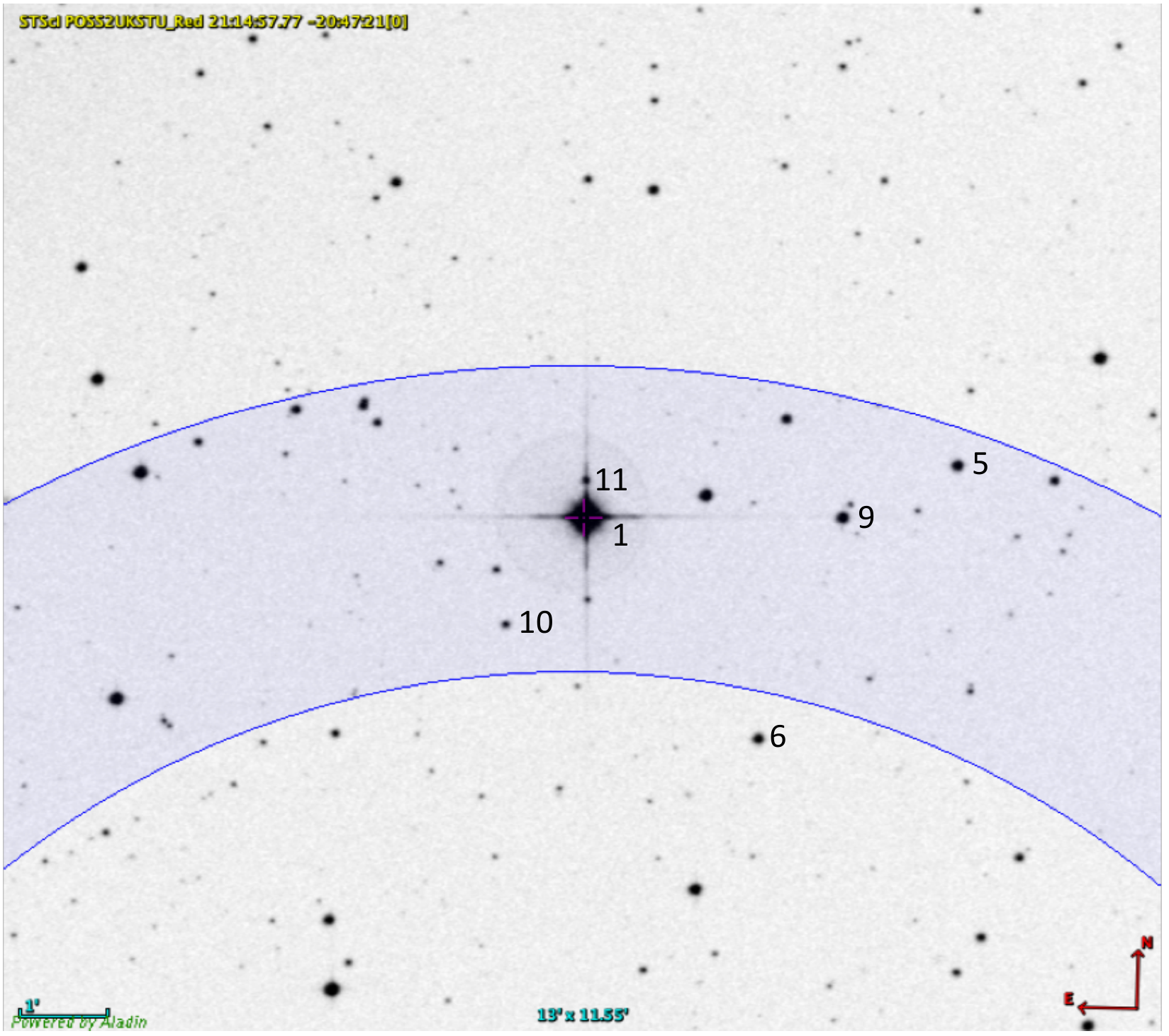}
\caption{Positions within FGS\,1r (blue-shaded region) of HD 202206 (1) and the astrometric reference stars (5 - 11) identified in Table~\ref{tbl-1}. Note that due to \HSTs roll restrictions, not all reference stars can be observed at each epoch. For example, ref-6 lies outside the FGS\,1r \ {FOV} at this epoch.}
\label{fig-Find}
\end{figure}
\end{center}

\begin{center}
\begin{figure}
\includegraphics[width=6in]{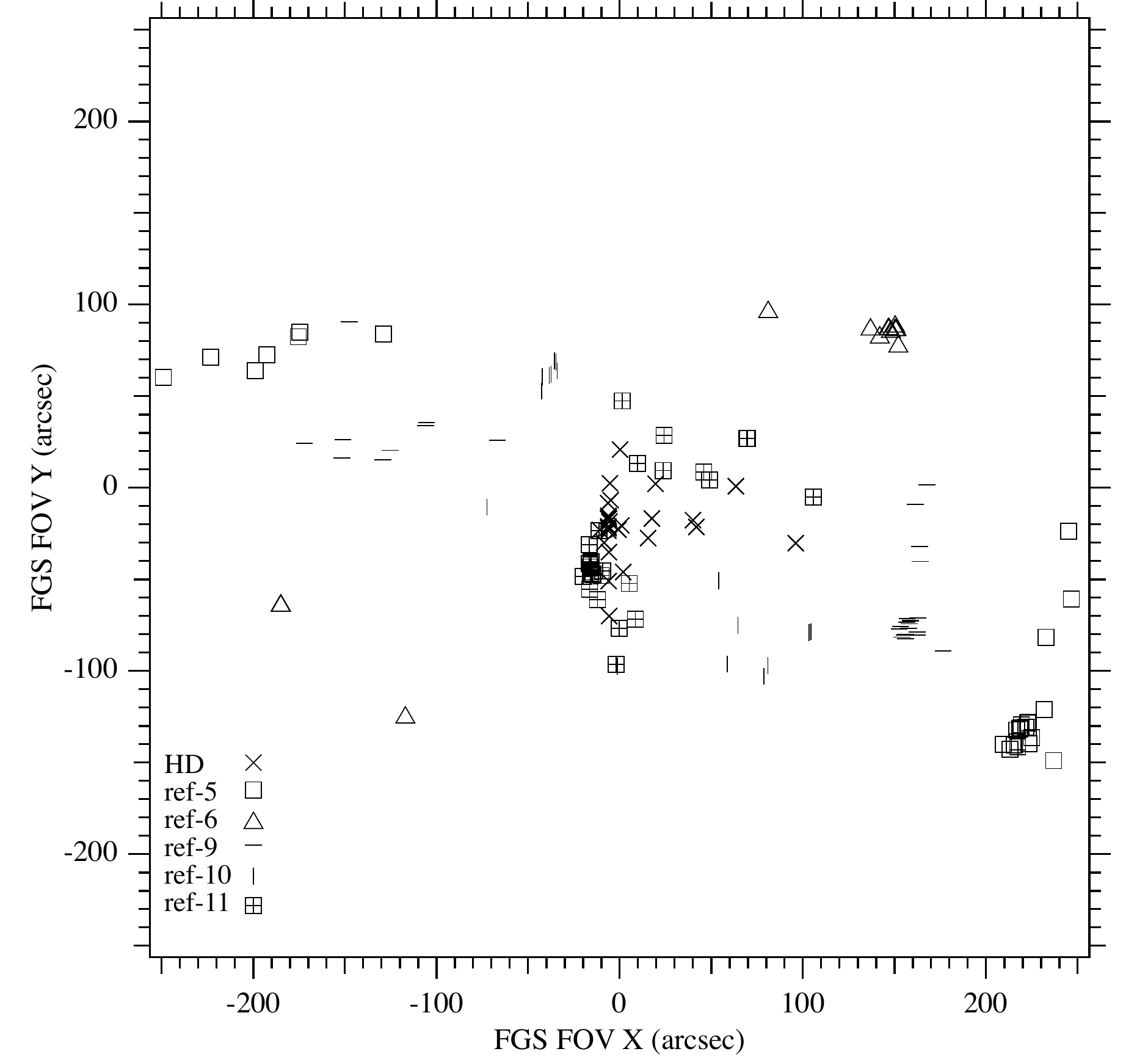}
\caption{Positions of HD 202206 (HD) and astrometric reference stars (5 --11) in FGS\,1r \ {FOV} coordinates. Due to guide star availability it was not possible to keep HD 202206 in the \ {FOV} center at each epoch, \ {but the distance from the FOV center always remained $\leq 100$ seconds of arc, with an average distance, $<r>=32"$.}}
\label{fig-Pick}
\end{figure}
\end{center}

The astrometric reference frame for \HDs consists of five stars (Table~\ref{tbl-1}). 
The \HDs field (Figure~\ref{fig-Find}) exhibits the distribution of  astrometric reference stars (ref-5 through ref-11) used in this study. 
The \HDs field was observed at a very limited range of spacecraft roll 
values (Table~\ref{tbl-DATA}). 
Figure \ref{fig-Pick} shows the distribution in FGS\,1r  coordinates of the thirty-one
 sets (epochs) of \HDs and  reference star measurements.    \HDs (labeled \textbf{`$\times$'}) had to be placed in many different locations within the FGS\,1r \  {total field of view (FOV)}
to maximize the number of astrometric reference stars in the FGS field of view and to insure guide star availability for the other two FGS units. \ {However, because the average radial distance of \HDs from FGS FOV center was $<r>=32"$, the astrometric impact of this displacement is indistinguishable from measurement noise.} At each epoch we measured each reference stars 1 -- 4 
times, and \HDs 3--5 times. 

 
\subsubsection{Modeling Priors}\label{CORR}
\begin{center}
\begin{figure}
\includegraphics[width=6in]{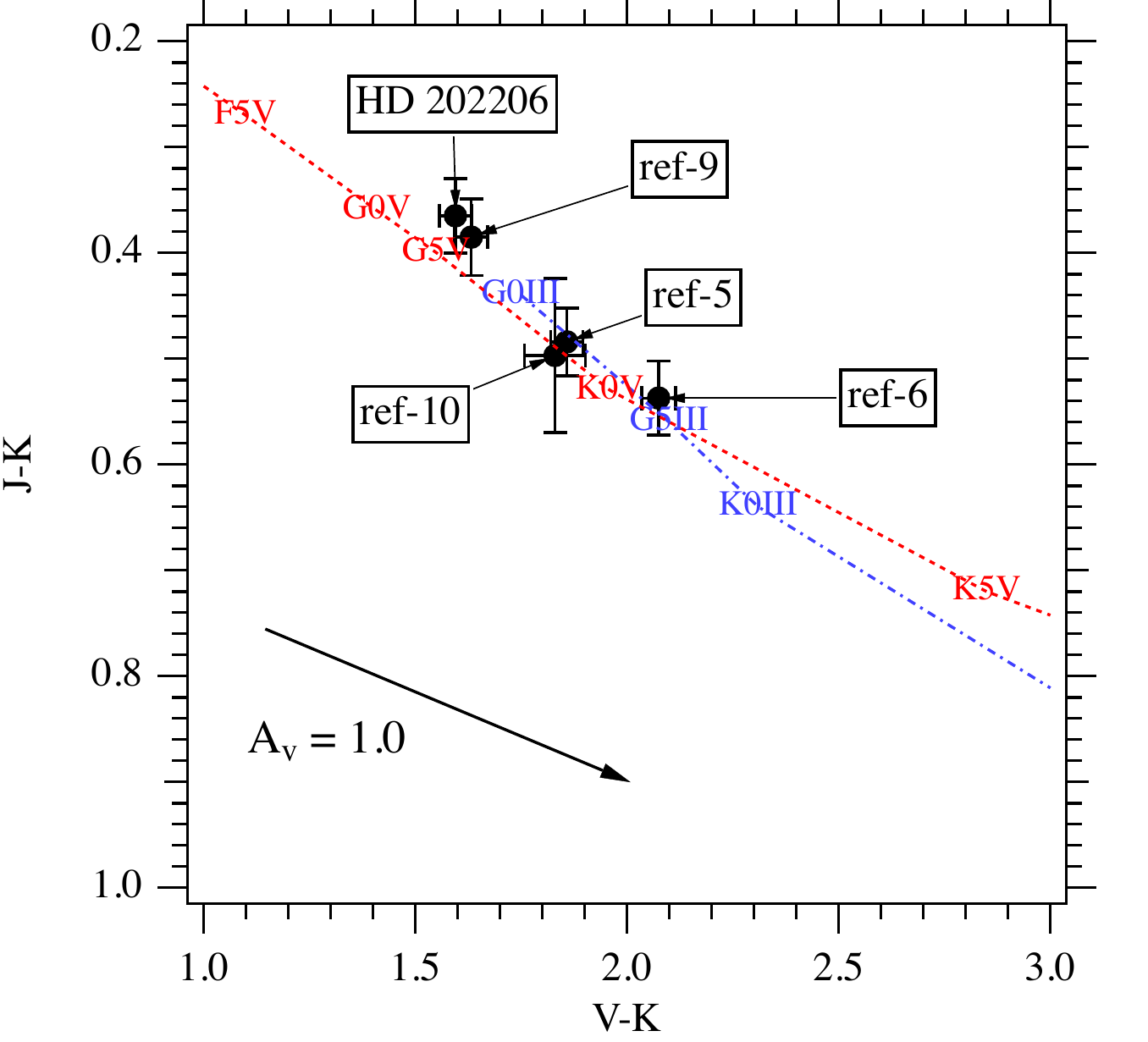}
\caption{$(J-K)$ vs. $(V-K)$ color-color diagram for HD 202206 and astrometric reference stars identified in 
Table~\ref{tbl-IR}. The  lines are the assumed \citep{Cox00} locus of  dwarf
(luminosity class V, dashed) and giant (luminosity class III, dot-dashed) stars of various spectral types. The reddening vector indicates $A_V = 1.0$ for the 
plotted color systems. Along this line of sight maximum extinction is  $A_V$$\sim$ 0.15 
\citep{Schl98}.  {There exists no 2MASS photometry for ref-11.}}
\label{fig-CCD}
\end{figure}
\end{center}

\begin{center}
\begin{figure}
\includegraphics[width=5in]{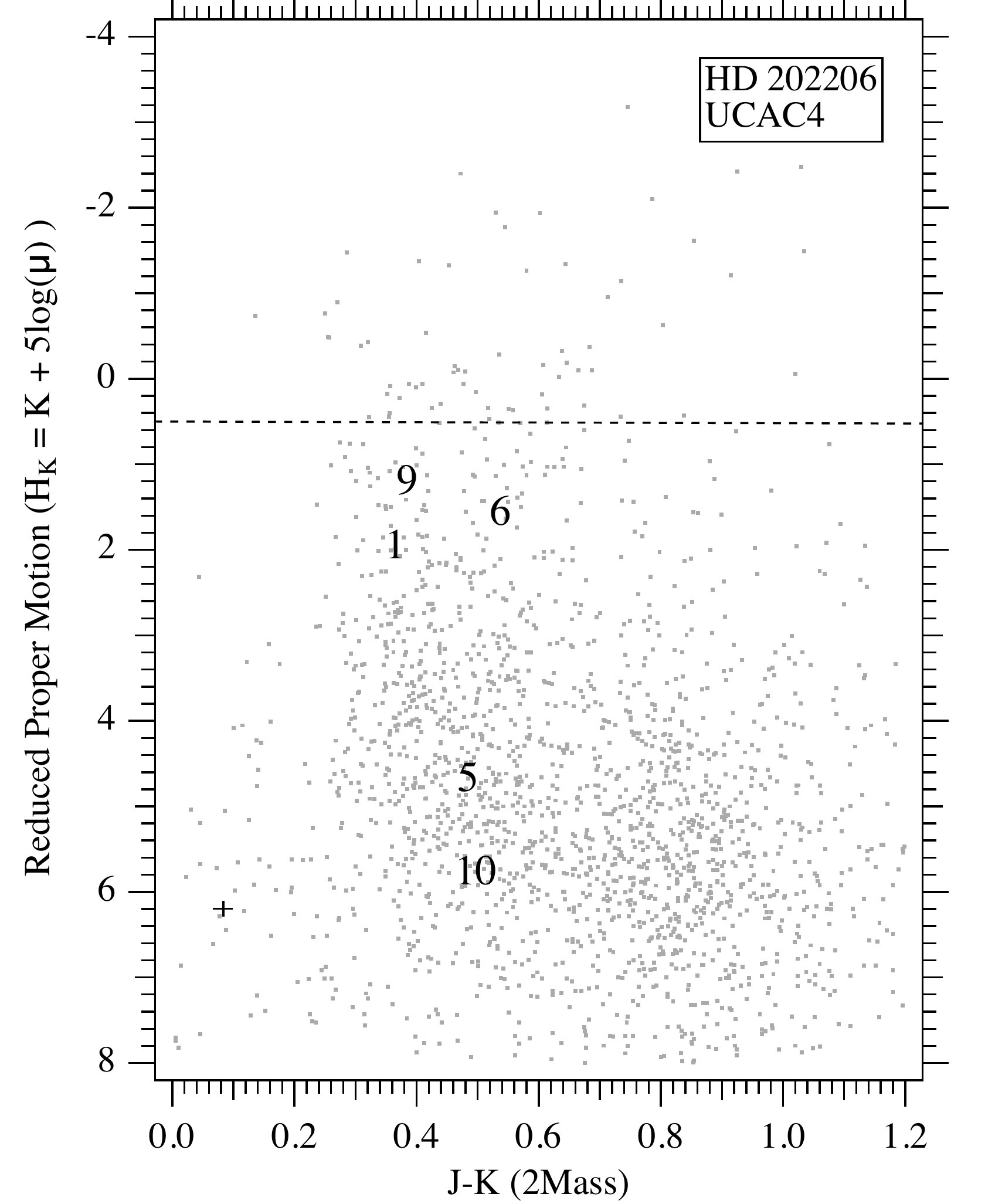}
\caption{Reduced proper motion diagram for 2200 stars in a 1\arcdeg ~field centered
on HD 202206.  Star identifications are in Table~\ref{tbl-IR} and proper motions are from Tables~\ref{tbl-PM} and \ref{tbl-SUM}. 
For a given spectral type, 
giants and sub-giants have more negative $H_K$ values and are redder than dwarfs in 
$(J-K)$.  $H_K$ values are derived from  proper motions in Table \ref{tbl-PM}. The small cross at the lower left represents a typical $(J-K)$ error of 0.04 mag and $H_K$ error of 0.17 mag.  Ref-11, omitted from the plot, lacks 2MASS photometry. \ {The horizontal line indicates a separation between dwarfs and sub-giant/giant stars.}
} 
\label{fig-RPM}
\end{figure}
\end{center}

The success of single-field parallax astrometry depends on prior knowledge of the reference stars, and sometimes, of the science target. Catalog proper motions with associated errors, lateral color corrections, and estimates for reference star parallax are entered into the modeling as quasi-Bayesian priors, data with which to inform the final solved-for parameters. These values are not entered as hardwired quantities known to infinite precision. We include them as observations with associated errors. The model adjusts the corresponding parameter values within limits defined by the data input errors to minimize $\chi^2$, yielding the most accurate parallax and proper motion for the prime target, \HD, and the best opportunity to measure any reflex motion due to the companions detected by RV.
\begin{enumerate} 
\item \textbf{Reference Star Absolute Parallaxes-} Because we measure the parallax of \HDs  with respect to 
reference stars which have their own parallaxes, we must either apply a statistically-derived correction from relative to absolute parallax \citep[Yale Parallax Catalog, YPC95]{WvA95}, or 
estimate the absolute parallaxes of the reference frame stars. We, again, choose the second option as we have since we first used it in \cite{Har99}. 
The colors, spectral type, and luminosity class of a star can be used to estimate the 
absolute magnitude, $M_V$, and $V$-band absorption, $A_V$. We estimate the absolute parallax for each reference star through this expression,
\begin{equation}
\pi_{\rm abs} = 10^{-(V-M_V+5-A_V)/5}
\end{equation}

Our band passes for reference star photometry include: $BVRI$ photometry of the reference stars from the NMSU 1 m
telescope located at Apache Point Observatory and JHK (from 2MASS\footnote{The Two Micron All Sky Survey
is a joint project of the University of Massachusetts and the Infrared Processing
and Analysis Center/California Institute of Technology }). Table \ref{tbl-IR} lists the visible and infrared photometry for the \HDs reference stars.

To establish spectral type and luminosity class, the reference frame stars were observed on 2009 December 9 using the
RCSPEC on the Blanco 4 m telescope at CTIO. We used the KPGL1 grating
to give a dispersion of 0.95 \AA /pix. Classifications used a combination of template matching and line ratios.  We determine the spectral types for the higher S/N stars to within $\pm$1 subclass. Classifications for the lower S/N stars have $\pm$2
subclass uncertainty. Table \ref{tbl-SPP} lists the spectral types and luminosity classes for our reference stars. Note that we had no prior IR photometry or spectral information for reference star, ref-11 (just above and quite close to \HDs in Figure~\ref{fig-Find}), hence no input prior parallax for the modeling.

Figure 
\ref{fig-CCD} contains a $(J-K)$ vs. $(V-K)$ color-color diagram for \HDs and the  reference stars. Schlegel \etal (1998)  find an upper limit $A_V$$\sim$0.15 towards 
\HD, consistent with the small absorptions we infer comparing spectra and photometry (Table~\ref{tbl-SPP}).
The reference star derived absolute magnitudes critically depend on the assumed stellar 
luminosity, a parameter impossible to obtain for all but the latest type stars 
using only  Figure \ref{fig-CCD}. To check the luminosity classes obtained from classification spectra we obtain  proper 
motions from the 
UCAC4 \citep{Zac13} for a one-degree-square field centered on \HD, and then 
produce a reduced proper motion diagram \citep{Str39,Yon03,Gou03}  to discriminate 
between giants and dwarfs. Figure~\ref{fig-RPM} contains the reduced proper motion diagram for the \HDs field, including \HDs and our reference stars. We derive absolute parallaxes by comparing our estimated spectral types and luminosity 
class to  $M_V$ values from \nocite{Cox00} Cox (2000). 

We adopted 1.0 mag input errors for 
distance moduli, $(m-M)_0$, for all reference stars. Contributions to the error are uncertainties in $A_V$ and errors 
in $M_V$ due to uncertainties in color to spectral type mapping.  We list all reference star absolute parallax estimates in Table \ref{tbl-SPP}.  Individually, no reference star absolute parallax is better determined than ${\sigma_{\pi}\over \pi}$ = 
23\%. The average input absolute parallax for the reference frame is 
$\langle\pi_{abs}\rangle = 1.7$ mas. We compare this to the correction to absolute parallax discussed and presented in 
YPC95 (section 3.2, figure 2).  Entering YPC95, figure 2, with the Galactic latitude of \HDs, $b = 
-40\arcdeg$, and average magnitude for the reference frame, $\langle V_{\rm ref} \rangle 
= 14.94$, we obtain a correction to absolute of 1.5 mas, consistent with our derived correction.

\item \textbf{Proper Motions-} We use proper motion priors from the UCAC4 Catalog \citep{Zac13}. These quantities typically have errors on order 
4 mas yr$^{-1}$.

\item \textbf{Lateral Color Corrections-}  \ {To effectively periscope the
entire FGS FOV, the FGS design includes refractive optics. Hence, a blue star and a red star at exactly the same position on the sky would be measured to have different positions. A series of observations of pairs of red and blue stars with small angular separation at various spacecraft roll positions yields the required corrections. The discussion in section 3.4 of \cite{Ben99} describes how we derive this correction for FGS\,3. A similar 
analysis resulted in FGS\,1r lateral color corrections $lc_x = -0.83 \pm 0.11$ mas and $lc_y = -0.8 \pm 0.08$ mas, quantities introduced as observations with error in the model shown below. These corrections have very little impact on the final results, given the small spread in $B-V$ color (Table~\ref{tbl-IR}) between \HDs and the reference stars.}
\end{enumerate}

\subsection{The Astrometric Model}
While the \HDs usable reference frame contains five stars, due to guide star availability we average four observed reference stars stars per epoch. From  
positional measurements we determine the scale, rotation, and offset ``plate
constants" relative to an arbitrarily adopted constraint epoch for
each observation set. We employ GaussFit (Jefferys \etal 
1988) 
\nocite{Jef88} to minimize $\chi^2$. The solved equations of condition for the 
\HDs 
field are:
\begin{equation}
        x^\prime = x + lc_x(\it B-V)  
\end{equation}
\begin{equation}
        y^\prime = y + lc_y(\it B-V) 
\end{equation}
 {
\begin{equation}
\xi = \overline{A}x^\prime + \overline{B}y' + \overline{C}  - \mu_\alpha \Delta t  - P_\alpha\pi- (ORBIT_{B,x} + ORBIT_{c,x})
\end{equation}
\begin{equation}
\eta = -\overline{B}x^\prime + \overline{A}y^\prime + \overline{F}  - \mu_\delta \Delta t  - P_\delta\pi- (ORBIT_{B,y} + ORBIT_{c,y})
\end{equation}
}
\noindent 
Identifying terms, $\it x$ and $\it y$ are the measured coordinates from {\it HST};   $(B-V)$ is the Johnson $(B-V)$ color of each star; and $\it lc_x$ and $\it lc_y$ are the lateral color corrections, which have little impact due to the small range of color for \HDs and reference stars (Table~\ref{tbl-IR}).  \ {$\overline{A}$, and $\overline{B}$ are scale and rotation plate constants, $\overline{C}$ and $\overline{F}$ are offsets}; $\mu_\alpha$ and $\mu_\delta$ are proper motions; $\Delta t$ is the time difference from the constraint epoch; $P_\alpha$ and $P_\delta$ are parallax factors;  and $\it \pi$ is  the parallax.   Note that we apply no cross-filter corrections (c.f. Benedict et al. 2007\nocite{Ben07}) because \HDs is faint enough that the FGS\,1r F5ND neutral density filter is unnecessary. 

We obtain the parallax factors from a JPL Earth orbit predictor 
(Standish 1990)\nocite{Sta90}, version DE405. We obtain an orientation to the sky for the 
FGS\,1r constraint plate (set 11 in Table~\ref{tbl-DATA}) from ground-based astrometry (the UCAC4 Catalog) with uncertainties 
of $0\fdg06$. 

$ORBIT_x$ and $ORBIT_y$ are functions of the classic parameters $\alpha$, the perturbation semi major axis, $i$, inclination, $e$, eccentricity, $\omega$, argument of periastron, $\Omega$, longitude of ascending node, $P$, orbital period, and $T_0$, time of periastron passage \citep{Hei78,Mar10}.
We model a sequence of measures of the host star motion (including parallax, proper motion and perturbations) relative to the reference frame seen in Figure~\ref{fig-Find}. 
 
The elliptical rectangular coordinates $x$,$y$, of the unit orbit are 
\begin{eqnarray}
x & = & (\cos{E} - e) \label{eq:xo}\\
y &  = & \sqrt{1-e^{2}} \sin{E} \label{eq:yo}
\end{eqnarray}

\noindent with eccentricity, $e$,  and $E$, the eccentric anomaly. $E$ depends on time, $t$,  through Kepler's equation,

\begin{equation}
\frac{2\pi}{P}(t - T_{0}) = E - e\sin{E} \label{Kep}
\end{equation}   

\noindent with epoch of periastron passage, $T_{0}$, and the orbital period, $P$. The eccentric anomaly, $E$ relates to the true anomaly, $f$, through
 
 \begin{equation}
\tan{\frac{f}{2}} = \sqrt{\frac{1+e}{1-e}} \tan{\frac{E}{2}}
\end{equation}

\noindent The projection of this true orbit onto a plane tangent to the sky yields the coordinates $ORBIT_x$, $ORBIT_y$

\begin{eqnarray}			  	  
ORBIT_ x & = & Bx + Gy \label{eq:x}\\
ORBIT_ y & = & Ax + Fy \label{eq:y}
\end{eqnarray}

\noindent with Thiele-Innes constants; \ {$B_{TI},A_{TI},G_{TI},F_{TI}$}

\begin{eqnarray}			  	  
B_{TI}& = & \alpha (\cos{\omega} \sin{\Omega} + \sin{\omega}\cos{\Omega}  \cos{i}) \\
A_{TI} & = & \alpha (\cos{\omega}\cos{\Omega}  -\sin{\omega} \sin{\Omega} \cos{i}) \\
G_{TI} & = & \alpha (- \sin{\omega}\sin{\Omega} +\cos{\omega} \cos{\Omega} \cos{i}) \\
F_{TI} & = & \alpha (-\sin{\omega}\cos{\Omega}  - \cos{\omega} \sin{\Omega} \cos{i}) 
\end{eqnarray}

\noindent  $ORBIT_x$ and $ORBIT_y$ denote the  coordinates of the parent star around the barycenter. For \HDs the FGS detects and characterizes a superposition of the perturbation sizes, $\alpha_B$ and $\alpha_c$ due to components B and c, through $(ORBIT_{B,x} + ORBIT_{c,x})$ and $(ORBIT_{B,y} + ORBIT_{c,y})$.


\subsection{The RV Model} \label{RV}

\cite{Udr02,Cor05,Cou10} measured the radial component of the stellar orbital motion around the barycenter of the system with Doppler spectroscopy. This changing velocity, $v$, is the projection of a Keplerian orbital velocity to the observer's line of sight plus a constant velocity $\gamma$. Therefore, for components B and c

\begin{eqnarray}
v_B = \gamma + K_B[\cos{(f_B+\omega_B)} + e_B \cos{\omega_B}]\\
v_c = \gamma + K_c[\cos{(f_B+\omega_c)} + e_c \cos{\omega_c}]\\
v_{tot} = v_B+v_c
\label{eq:rv}
\end{eqnarray}      

\noindent  where $K$ is the velocity semi-amplitude. The total RV signal \citep{Cou10} we model ($v_{tot}$) includes contributions from both components B and c.

\subsection{Determining Perturbation Orbits for  \HDs B and c}
To derive companion perturbation orbital elements we simultaneously model  RV values from \cite{Cou10} and \HSTs astrometry (Table~\ref{tbl-DATA}). 
Because our GaussFit modeling results critically depend on the input data errors, we first modeled only the RV (Equation \ref{eq:rv}) to assess the validity of the original \citep{Cou10} input RV errors.  Solving for the orbital parameters of components B and c, to achieve a $\chi^2/$DOF of unity, where DOF represents the degrees of freedom in the solution, required increasing the original errors by a factor of 1.4. 

Tables~\ref{tbl-PM} and \ref{tbl-SUM} list results of this modeling; the proper motions (relative), absolute parallaxes, and absolute magnitudes and their errors (1-$\sigma$) for the five reference stars and \HD. 
Table~\ref{tbl-ORB} contains final orbit parameter values and errors for a model including both RV and astrometry; the period ($P$), the epoch of passage through periastron in years ($T$), the
eccentricity ($e$), and the angle in the plane of the true orbit between
the line of nodes and the major axis ($\omega$), are  the same for an orbit determined from
RV or from astrometry.  
The remaining orbital elements ($i, \Omega, \alpha$) come only from astrometry. Our model allows  the astrometry and the RV to describe two companions, \HDs B and c. Astrometry and RV are forced to describe the same system
through this constraint \citep{Pou00}, shown for component B, though in the model applied to both the B and c components,
\begin{equation}
\displaystyle{{\alpha_B~sin\,i_B \over \pi_{abs}} = {P_B K_B (1 -
e_B^2)^{1/2}\over2\pi\times4.7405}} 
\label{PJ}
\end{equation}
\noindent where quantities derived only from astrometry (parallax, $\pi_{abs}$, 
host star perturbation orbit size, $\alpha$, and inclination, $i$) are on the left, and 
quantities  derivable from both (the period, $P$ and eccentricity, $e$), or radial 
velocities only (the RV amplitude of the primary, $K$, induced by a companion), are on the right. Given the sparse orbit coverage of the \HDs B and especially the c perturbation afforded by the astrometry (Figures~\ref{Fig-ORB} and~\ref{Fig-ORBT}),  the RV data were essential in determining the component orbits. For most of the orbital parameters in Table~\ref{tbl-ORB} a combination of astrometry and previously existing RV has reduced the \cite{Cor05,Cou10} formal errors.

\subsection{Assessing Modeling Residuals}
\begin{figure}
\includegraphics[width=4in]{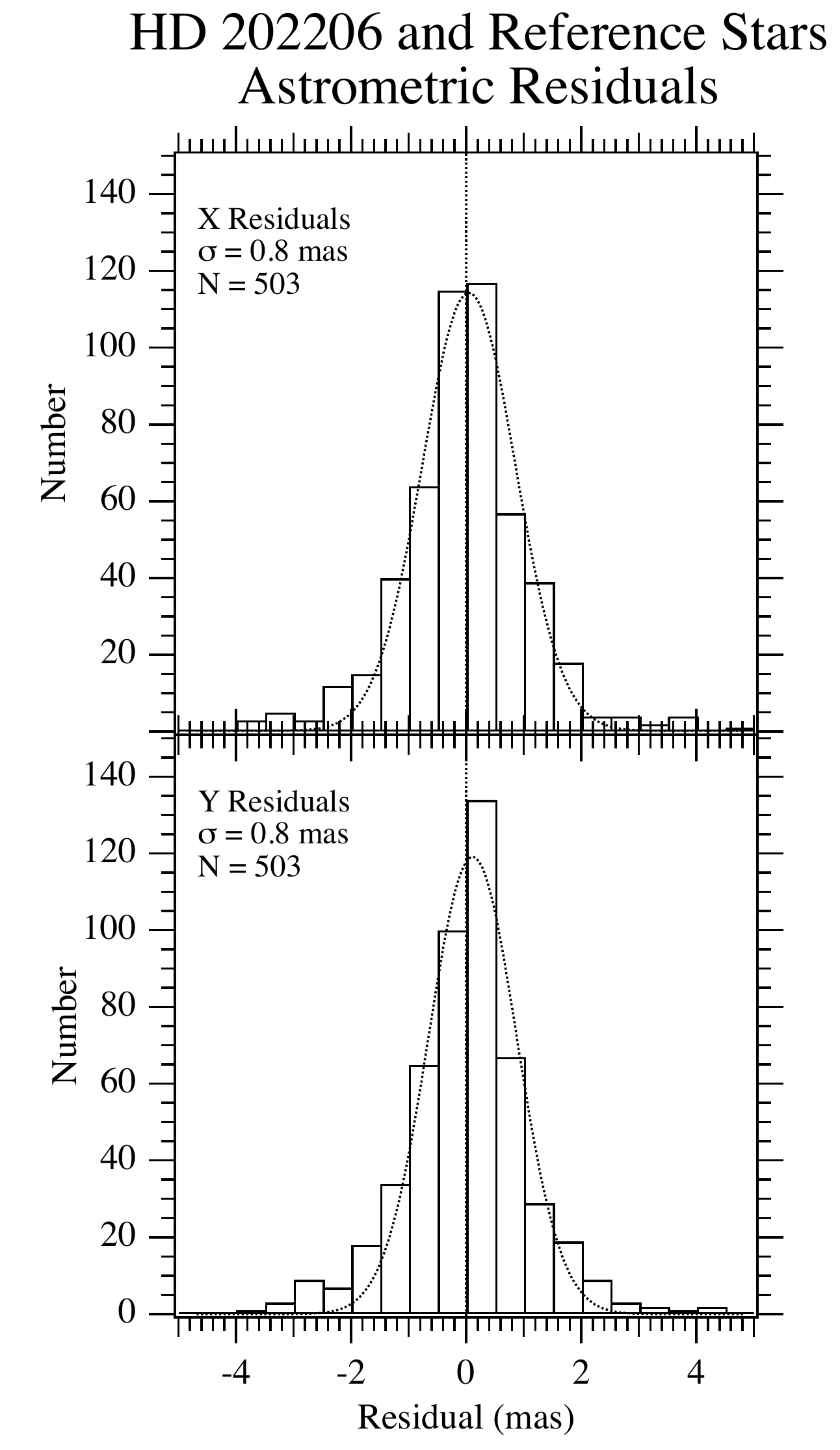}
\caption{Histograms of x and y residuals obtained from modeling the FGS 
observations of \HDs and the FGS reference frame with Equations 2 -- 5. Distributions are 
fit with gaussians with standard deviations, $\sigma$, indicated in each panel.} \label{fig-FGSH}
\end{figure}

From 
histograms of the FGS astrometric residuals (Figure~\ref{fig-FGSH}) we conclude 
that we have a well-behaved solution exhibiting residuals with Gaussian distributions with dispersions $\sigma \sim 0.8$ mas. \ {The slight skew in the Y residuals can be seen in either X or Y residuals, either positive or negative in many previous modelings, e.g. 
\cite{Ben09,Ben10,Ben11,McA11, Ben16} with no discernable impact on results.} The  reference frame 
'catalog' from  FGS\,1r in $\xi$ and $\eta$ standard coordinates (Table \ref{tbl-POS}) 
was determined with average uncertainties, $\langle\sigma_\xi \rangle= 0.26$ and $\langle\sigma_\eta \rangle = 0.22$ mas. Because we have rotated our constraint plate to an RA, DEC coordinate system, $\xi$ and $\eta$ are RA and DEC.

\begin{figure}
\includegraphics[width=6.5in]{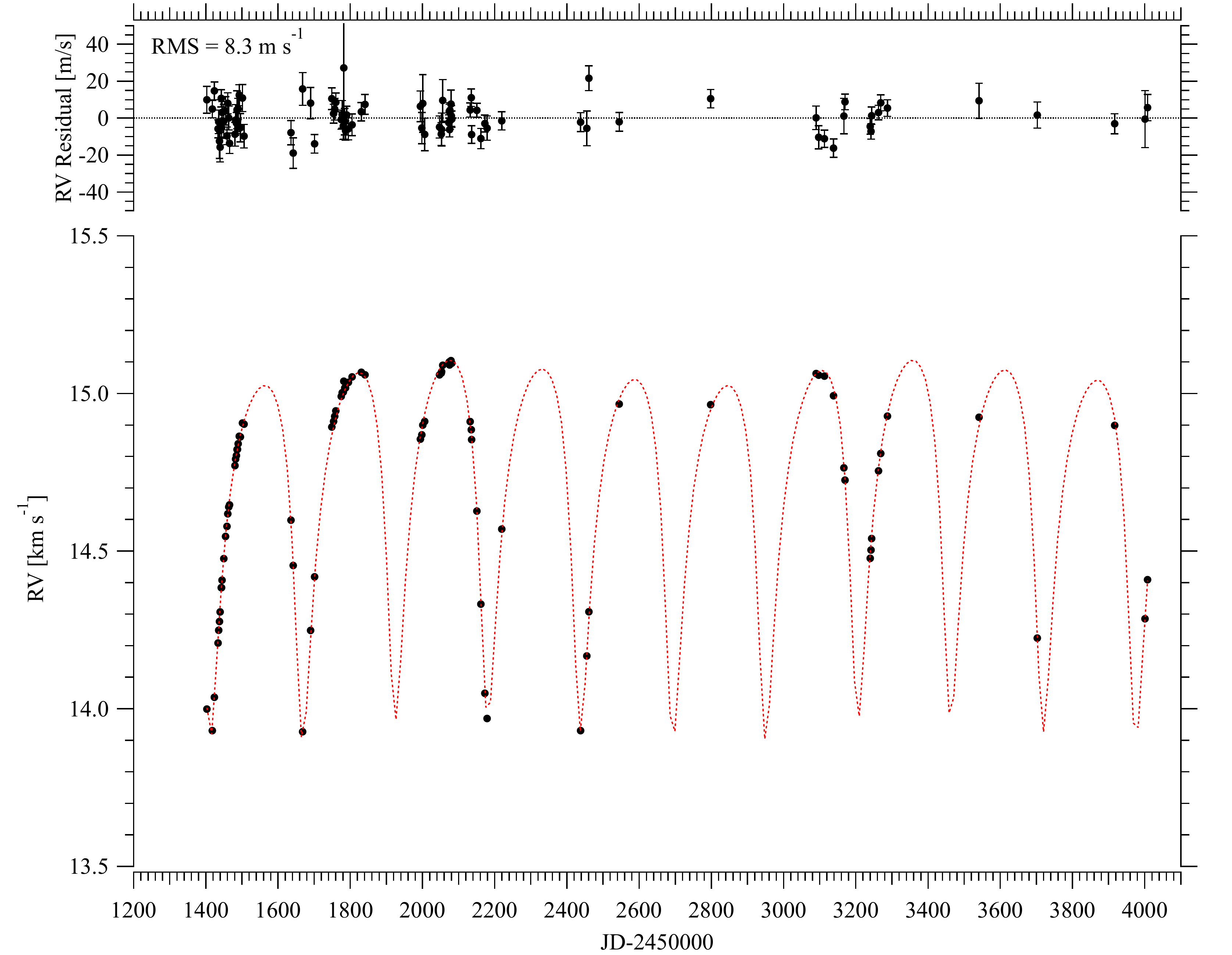}
\caption{RV values from \cite{Cou10} and the final RV two component orbit (Table~\ref{tbl-ORB}) obtained from modeling the RV and the FGS 
observations of \HDs and the FGS reference frame with Equations 2 -- 5. The original RV errors \citep{Cou10} have been increased by a factor of 1.4 to achieve a unity $\chi^2$. Residuals are plotted in the top panel
tagged with the adopted RV errors. We note the RMS residual value in the plot.} \label{fig-RVf}
\end{figure}

\begin{figure}
\includegraphics[width=6.5in]{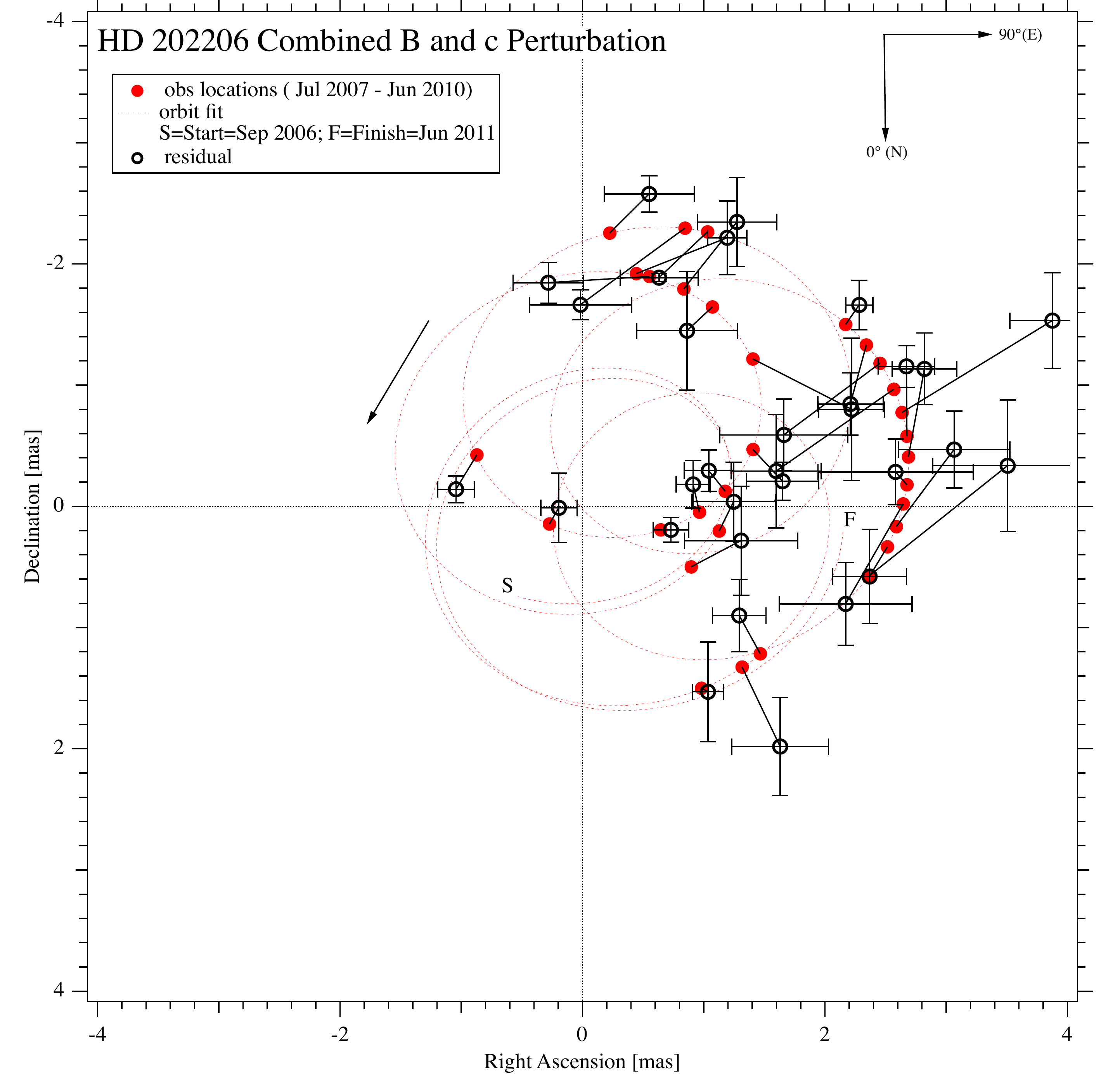}
\caption{Residuals to the combined B,c perturbation described by the Table~\ref{tbl-ORB} final orbital elements. Normal points (o) for each Table~\ref{tbl-DATA} epoch (unique set number) attach to their calculated locations ($ \bullet $) on the combined orbit (- -), representing 4.8 years from September 2006 to July 2011.  Actual observations spanned July 2007 to June 2010. Residual RMS is 0.35 mas in RA, 0.32 mas in DEC. Errors are the standard deviation of the mean for each normal point, typically comprised of 5 separate observations per set.\label{Fig-ORB}}
\end{figure}

\begin{figure}
\includegraphics[width=6.5in]{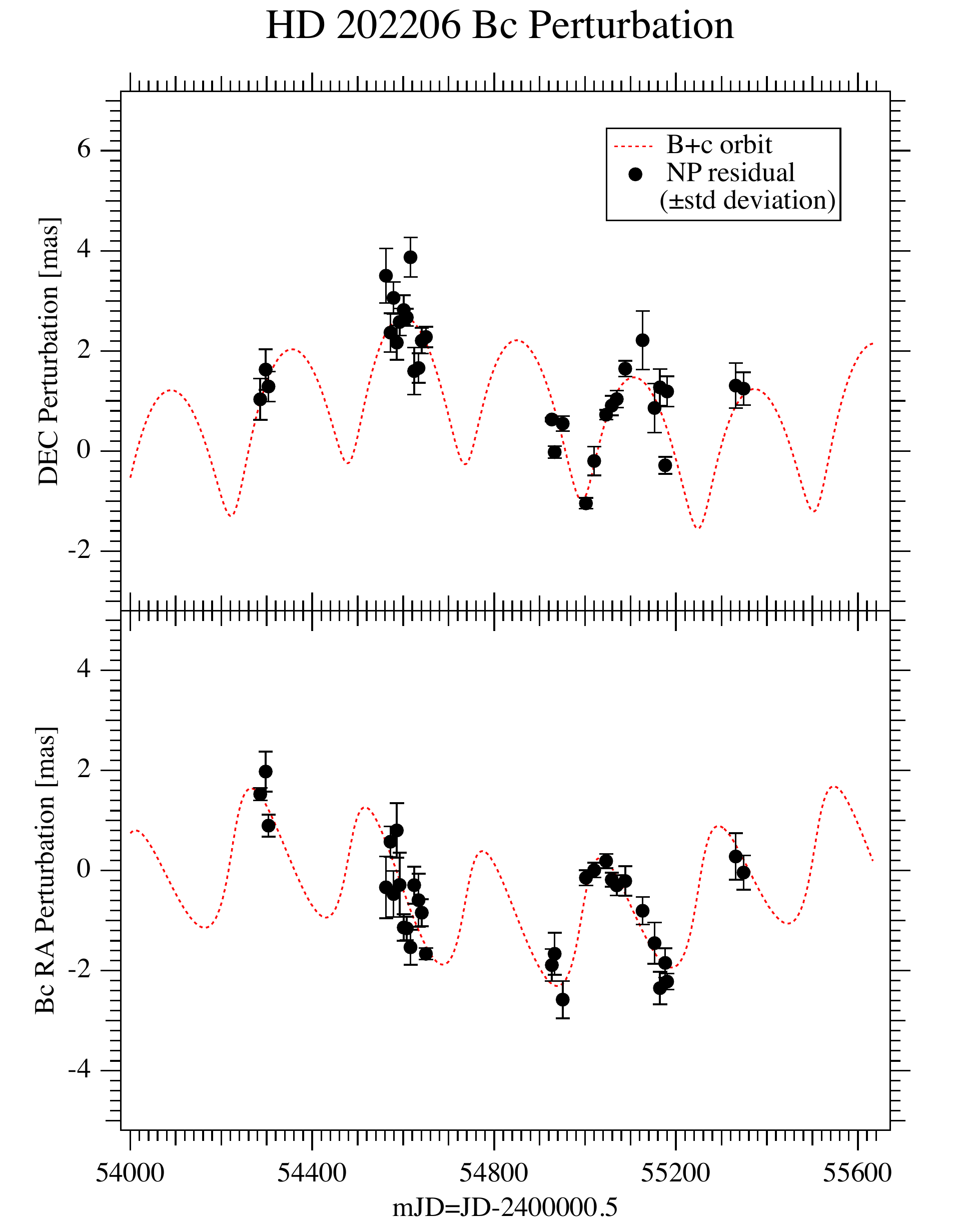}
\caption{Time variation of the RA and DEC components of the perturbation (- -) described by the Table~\ref{tbl-ORB} final orbital elements over-plotted with normal points ($ \bullet $) for each Table~\ref{tbl-DATA} epoch (unique set number).   Errors are the same as for Figure~\ref{Fig-ORB}. \label{Fig-ORBT}}
\end{figure}

At this stage we can assess the quality of the \HDs B and  \HDs c astrometric perturbations by plotting 
the RV and astrometric residuals from our modeling of the component B,c orbit. We show the RV orbit with adopted errors and final residuals to the simultaneous modeling in Figure~\ref{fig-RVf}.
Figure~\ref{Fig-ORB} shows the RA and DEC 
components at each observational epoch (the 31 data sets listed in Table~\ref{tbl-DATA}) plotted on the final component B,c orbit. We plot averages of FGS residuals at each epoch plotted as small symbols, connected to their calculated position on the orbit.   These normal point residuals have an average absolute value residual, $\langle |residual| \rangle 
= 0.34$ mas. Figure~\ref{Fig-ORBT} shows our average (typically five positions) measures for each Table~\ref{tbl-DATA} data set with associated standard deviation of the mean plotted on the RA and DEC components of the combined B,c orbit described by the model-derived orbital elements  in Table~\ref{tbl-ORB}. 

\section{Masses and Mutual Inclination} \label{MMI}
For the parameters critical in determining 
the masses of the companions to \HDs we find a parallax, $\pi_{abs} = 21.96 \pm 0.12$ mas and a proper 
motion in RA of  $-41.54\pm 0.11$  mas y$^{-1}$ and in DEC of $-117.87 \pm 0.11$ mas y$^{-1}$. 
Table~\ref{tbl-SUM}  compares values for the parallax and proper motion of \HD 
~from {\it HST}, \G~\citep{Bro16}, and  the   \HIP~re-reduction \citep{Lee07a}. While the parallax values agree within their respective errors, we note a small disagreement in the  proper motion vector ($\vec{\mu}$) absolute magnitude and direction. This could be explained by our non-global proper motion measured against a small sample of reference stars. 
Our measurement precision and extended study duration have significantly improved the precision of the parallax of \HD.

For the perturbation due to component B we find $\alpha_B = 1.4 \pm 0.1$ mas, and an inclination, 
$i_B $= 10\fdg9 $\pm$ 0\fdg8. 
We find $\alpha_c = 0.76 \pm 0.11$ mas, and an inclination, 
$i_c $= 7\fdg7 $\pm$ 1\fdg1. We list all modeled orbital elements in Table~\ref{tbl-ORB} with 1-$\sigma$ errors. The mutual inclination, $\Phi$, of the B and c orbits can be determined through \citep{Kop59, Mut06}
\beq
cos\,\Phi = cos\,i_B cos\,i_c + sin\,i_B sin\,i_c cos(\Omega_B - \Omega_c) \label{MI}
\eeq

\noindent where $i_B$ and $i_c$ are the orbital inclinations and $\Omega_B$ and  $\Omega_c$  are the longitudes of their ascending nodes.
Our modeling yields a suggestion of coplanarity with $\Phi = 6 \pm 2 \arcdeg$.

\begin{figure}
\includegraphics[width=7in]{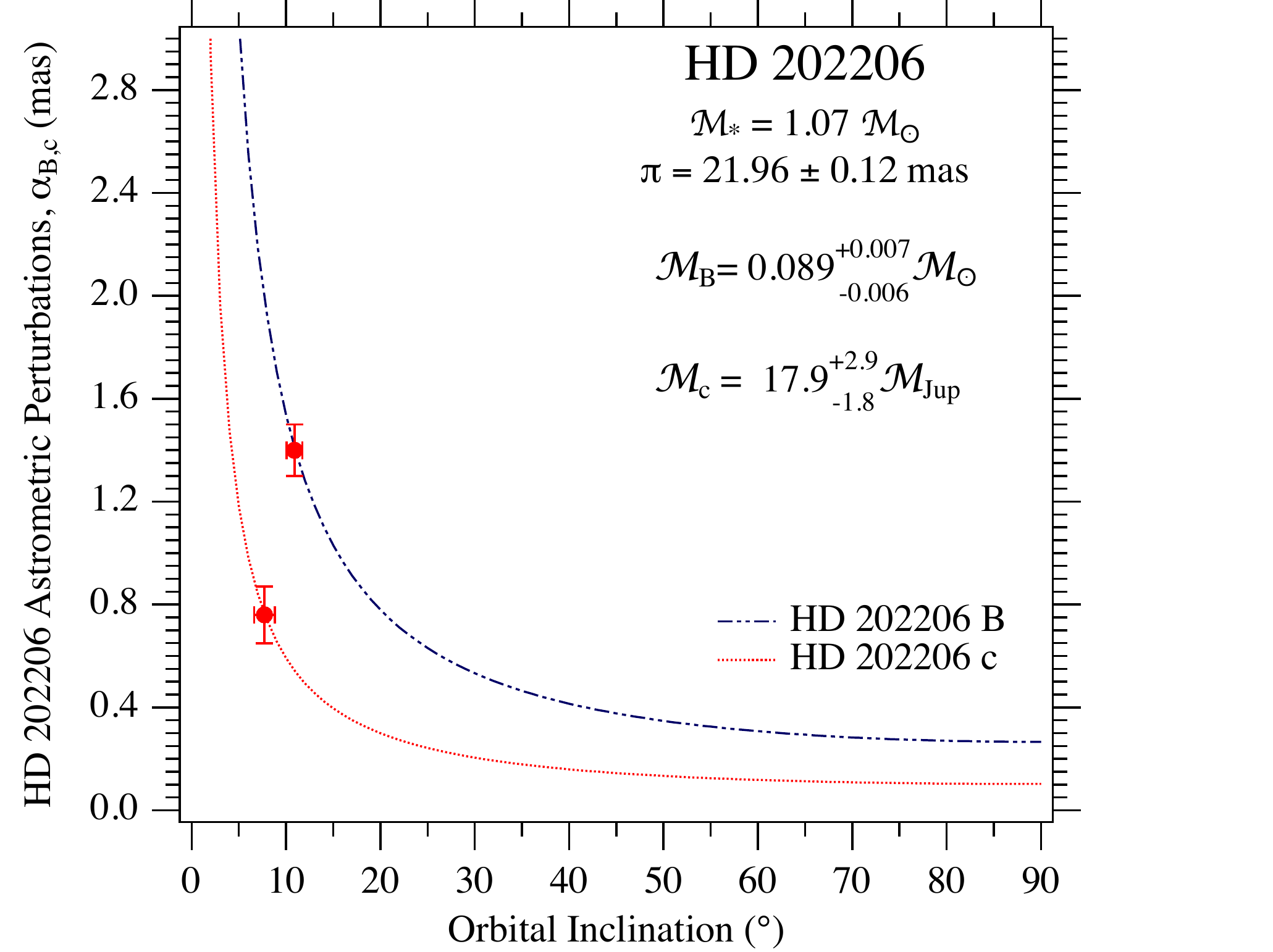}
\caption{These curves relate perturbation size and inclination for  \HD~B and c  through the Pourbaix \& Jorrisen (2000)  relation (Equation~\ref{PJ}). For each component we use the curve as a `prior' in a quasi-bayesian sense. Our final values for the semimajor axes of the astrometric perturbations, $\alpha_B$, $\alpha_c $ and inclinations, $i_B$,  $i_c$ are plotted with their formal errors. } \label{fig-perts}
\end{figure}

Figure~\ref{fig-perts} illustrates the Pourbaix and Jorrisen relation (Equation~\ref{PJ}) between parameters 
obtained from astrometry and RV and our final  
estimates for each component $\alpha$ and $i$. 
In essence, our simultaneous solution uses the 
Figure~\ref{fig-perts} component B and c curves as quasi-Bayesian priors, sliding along them until the 
astrometric residuals and orbit parameter errors are minimized. 

The planetary mass depends on the mass of the primary star, for which we have 
adopted \m$_*$=1.07 \msun \citep{Han14}. 
We find \m$_B = 93.6 ^{ +7.7}_{-6.6}$\mjupe ~= $0.089^{+0.007}_{-0.006}$ \msune. The central mass controlling the component c orbit is now the sum of the component A and B masses,  \m$_{A+B} = 1.16$\msune. Hence, for component c, \m$_c = 17.9^{+2.9}_{-1.8}$\mjupe. In Table~\ref{tbl-ORB} the final mass values for components B and c do not incorporate the present uncertainty 
in the stellar mass, \m$_*$. 

Table~\ref{tbl-SUM} shows the FGS proper motion to have a small disagreement with previously measured \HIP~ and \G~ values. Our modeling can include any priors, but we generally resist including priors for the prime scientific target. If we include \HDs proper motion priors (and estimated errors) from \HIP ~\citep{Lee07a}, \G ~\citep{Bro16}, the PPMXL \citep{Roe10}, UCAC4 \citep{Zac13}, and SPM 4.0 \citep{Gir11} catalogs, we obtain a proper motion in agreement with the \G~value. The resulting masses and mutual inclinations of components B and c agree within the Table~\ref{tbl-ORB} errors. However, the $\chi^2$ increases by 5.2\%, while the degrees of freedom increase by 1.1\%. Hence, we prefer the Table~\ref{tbl-ORB} results from a solution without \HDs proper motion priors.

\section{Discussion} \label{Disc}

From the \cite{Ben16} Mass-Luminosity relations we can estimate absolute magnitudes for an M dwarf star with mass \m = 0.089 \msune. 
Those relations yield $M_V=17.80$ and $M_K=9.79$. Our parallax,  $\pi_{abs}=21.96$, and an interstellar absorption, $A_V=0$, provide a distance modulus, 
$(m-M)_0 = 3.30$, and for the host star \HDs $M_V=4.77$ and $M_K=3.19$. \HDs B at apastron has a separation $\rho_B = a_B(1+e_B) = 26.1$ mas
with $\Delta V = 13.0$ and $\Delta K = 6.6$, a challenging upcoming (March-December 2019) test for any existing high-contrast imaging system.

Our characterization of the \HDs system comes close to providing a solution to the vexing problem of stability. With only \m \,sin\,i values for components b and c \cite{Cor05,Cou10,Pet15} argued that a stable \HDs system should be in a 5:1 mean motion resonance (MMR) and coplanar. Our re-determination of the periods (listed in Table~\ref{tbl-ORB}) yield $P_c/P_B = 4.92\pm0.04$, a value less than 3-$\sigma$ from MMR. Our mutual inclination, $\Phi = 6\pm2\arcdeg$, differs from coplanarity by 3-$\sigma$. 

Our modeling platform, GaussFit, easily accommodates any priors as data with associated errors. Given that stability seems to require a 5:1 MMR, we constructed a model that includes a new piece of `data', $Pdiff$, a new parameter, $Ppdiff$, and the associated equation of condition relating the two
\begin{eqnarray}
Pdiff = (P_B*5.0)-P_c\\
value =Ppdiff-Pdiff
\end{eqnarray}
\noindent This addition to our model introduces the period ratio, $P_c/P_B =5$, as a prior constraint, where the  observable derived from theory is $Pdiff=0\pm20$ days, and $value$ is the quantity to be minimized (in $\chi^2$) by the modeling. The adopted error for 
$Pdiff$ represents a 1.7\% difference in the expected 5:1 MMR. We present the orbital parameters and component masses resulting from that modeling in Table~\ref{tbl-resRES}, which now includes the parameter, $Ppdiff = 3 \pm 5$ days, effectively zero, suggesting a 5:1 MMR. The component masses are a little higher, but agree within the errors with those (Table~\ref{tbl-ORB}) resulting from a model with no prior knowledge of a possibly required resonance. The parallax and proper motions were unchanged from the Table~\ref{tbl-SUM} values. The assertion of a 5:1 MMR has forced a higher degree of coplanarity, i.e., the smaller $\Phi$ value shown in Table~\ref{tbl-resRES}.

\begin{figure}
\includegraphics[width=7in]{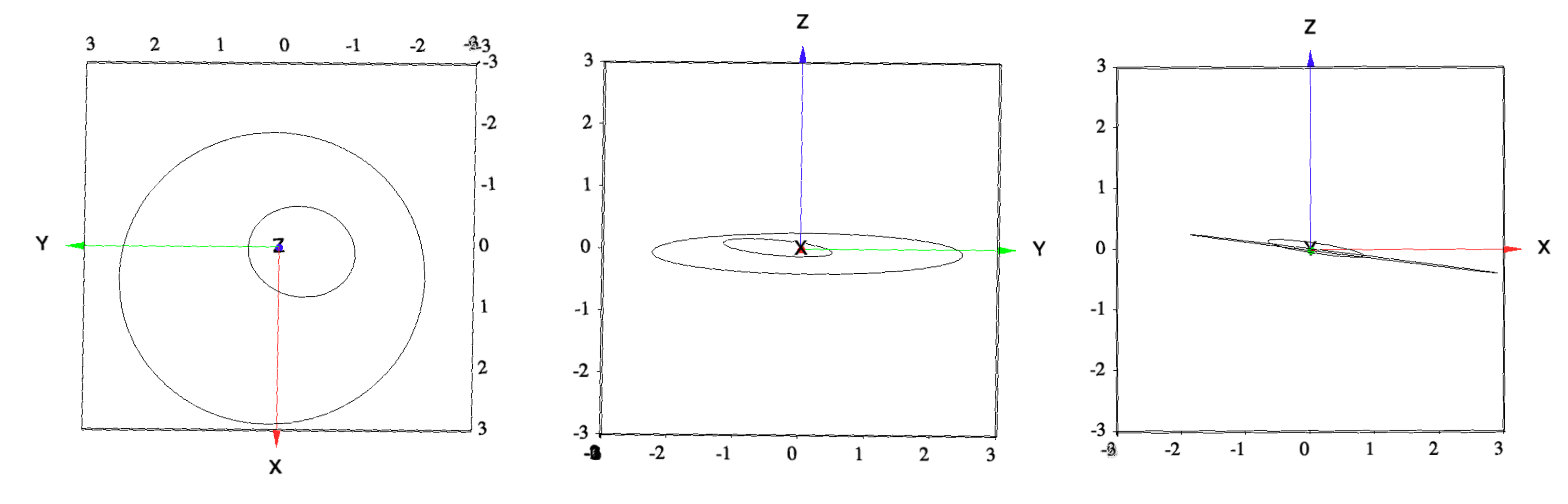}
\caption{Component B (inner) and c (outer) orbits as observed (left to right panels) towards the -z (looking at the \HDs system with +y to the south and +x pointing west), towards -x (east in RA), and -y (north in DEC) axes. Axes units are AU.} \label{fig-3D}
\end{figure}

Finally, we plot component B and c actual orbits (in AU, from the Table~\ref{tbl-ORB} parameters) in Figure~\ref{fig-3D} from three vantage points; as seen on the sky (along the z axis), and plots looking north towards -y  and east in the direction of -x. These views demonstrate the degree of coplanarity (without prior knowledge of a coplanarity requirement) determined through our modeling. 

\section{Summary} \label{Summ}

For  the \HDs system we find 
\begin{enumerate}
\item A parallax, $\pi_{abs}=21.96\pm0.12$ mas, agreeing with the \HIP~and $Gaia$ values within the errors,

\item A {\it relative, not absolute} proper motion relative to our reference frame, $\vec{\mu} = 124.98$ mas  yr$^{-1}$
with a position angle, P.A. = $199\fdg4$, differing by 1.5 mas yr$^{-1}$ and $1\fdg3$ compared to $Gaia$,

\item An inclination for \HD\,B , $i_B $= 10\fdg9 $\pm$ 0\fdg8 and, with the assumption of a \HD\,A mass, \m$_A$ = 1.07 \msun, a component B mass,  \m$_B = 0.089^{+0.007}_{-0.006}$ \msune.  \HD\,B is an \ {M8 dwarf star \citep{Dup17}},

\item A component c inclination, $i_c=7\fdg7 \pm 1\fdg1$, that with a central mass now \m$_{A+B}=1.16$ \msun, yields a component c   mass, \m$_c = 17.9^{+2.9}_{-1.8}$\mjupe. \HD\,c is a brown dwarf,
 
\item A period ratio $P_c/P_B = 4.92\pm0.04$, near a 5:1 MMR, and a flat \HDs system architecture with a B-c mutual inclination of $\Phi =6\arcdeg \pm 2\arcdeg$, near coplanarity,

\item That including proper motion priors from multiple sources yields the same component B and component c masses as ignoring those priors,

\item That including a 5:1 MMR as a prior yields the same component B and component c masses as ignoring that prior, while nudging the \HDs system slightly closer to coplanarity, with $\Phi =4\arcdeg \pm 2\arcdeg$.

\end{enumerate}

Thus the question posed in the title of the \cite{Cor05} paper, ``A pair of planets around HD 202206 or a circumbinary planet?", is answered with a single word; neither. The \HDs system consists of a low-inclination, nearly face-on G8V + M6V binary orbited by a brown dwarf. 

A combination of additional RV measurements and \G~astrometry should further illuminate our understanding of the dynamics of this interesting system, particularly by reducing the errors on periods and coplanarity. We repeat  an old question: is the \HDs system stable, or just close to stable?

\acknowledgments

Support for this work was provided by NASA through grants 11210 and 11788
 from the Space Telescope 
Science Institute, which is operated
by the Association of Universities for Research in Astronomy, Inc., under
NASA contract NAS5-26555. This publication makes use of data products from the 
Two Micron All Sky Survey, which is a joint project of the University of 
Massachusetts 
and the Infrared Processing and Analysis Center/California Institute of 
Technology, 
funded by NASA and the NSF.  This research has made use of the {\it SIMBAD} and {\it Vizier} databases, 
operated at Centre Donnees Stellaires, Strasbourg, France; Aladin, developed and maintained at CDS; the NASA/IPAC Extragalactic Database (NED) 
which is operated by JPL, California Institute of Technology, under contract 
with 
NASA;  and NASA's truly essential Astrophysics Data System Abstract Service. This work has made use of data from the European Space Agency (ESA)
mission {\it Gaia} (\url{https://www.cosmos.esa.int/gaia}), processed by
the {\it Gaia} Data Processing and Analysis Consortium (DPAC,
\url{https://www.cosmos.esa.int/web/gaia/dpac/consortium}). Funding
for the DPAC has been provided by national institutions, in particular
the institutions participating in the {\it Gaia} Multilateral Agreement. Many people over the years have materially improved all aspects of the work reported, particularly Linda Abramowicz-Reed, Art Bradley, Denise Taylor, and all the co-authors of our many papers. 
G.F.B. thanks Debbie Winegarten, whose able assistance with other matters freed me to devote necessary time to this analysis. 
Next, our thanks to Dr. Correia for the title of his 2005 paper, announcing a second companion. Nothing beats a question mark in a title for getting one's attention. Thanks to Barbara McArthur, whose analysis knowledge informed every aspect of this paper. Lastly, thanks to an anonymous referee, whose careful and critical reading resulted in suggestions that improved the final result.

\bibliography{/Active/myMaster}

\clearpage

\begin{deluxetable}{lll}
\tablewidth{0in}
\tablecaption{HD 202206 Stellar Parameters \label{tbl-STAR}}
\tablehead{ 
\colhead{Parameter}& 
\colhead{Value}& 
\colhead{Source}
}
\startdata
SpT&G6V&1\\
T$_{eff}$&5766 K&6\\
log g&4.5 $\pm$ 0.1&4\\
$[Fe/H]$&0.3 $\pm$ 0.1&6\\
age&2.9 $\pm$ 1.0 Gy&5\\
mass&1.07 $\pm$ 0.08 $\cal{M}_{\sun}$&4\\
distance&45.5 $\pm$ 0.3 pc & 2\\
A$_V$ & 0.0 & 1\\
Radius&1.04 $\pm$ 0.01$ R_{\sun}$&5\\
v\,sin\,i&2.3 $\pm$ 0.5 \kms&4\\
m-M&3.30$\pm$ 0.01&2\\
$V$&8.07 $\pm$ 0.01&1\\
$K$&6.49 $\pm$ 0.02&3\\
$V-K$&1.58 $\pm$ 0.03&1,3\\
\hline
\enddata
\\$^1$SIMBAD,$^4$Exoplanets Website \citep{Han14},
$^2$this paper, \\$^3$2MASS, $^5$\cite{Bon16}, $^6$\cite{Hin16}.
\end{deluxetable}

\begin{landscape}
\begin{deluxetable}{l l l r r r r r r r r r}
\tabletypesize{\tiny}
\tablewidth{0in}
\tablecaption{HD 202206 Field Astrometry\tablenotemark{a} \label{tbl-DATA}}
\tablehead{ 
\colhead{Set}& 
\colhead{Star}& 
\colhead{\HST ID} &
\colhead{V} &
\colhead{V3 roll} &
\colhead{X} &
\colhead{Y} &
\colhead{$\sigma_X$} &
\colhead{$\sigma_Y$} &
\colhead{t$_{\rm obs}$} &
\colhead{P$_{\alpha}$} &
\colhead{P$_{\delta}$}
}
\startdata
1&1&F9YM0102M&8.2&280.612&-5.3808719&2.5181351&0.0024&0.0030&54285.60839&0.579256385&0.098856994\\
1&5&F9YM0103M&14.41&280.612&232.7332958&-81.5462095&0.0035&0.0030&54285.60979&0.578137858&0.098769874\\
1&9&F9YM0105M&13.96&280.612&163.6020442&-32.1472399&0.0035&0.0038&54285.61266&0.578449648&0.098684625\\
1&5&F9YM0106M&14.42&280.612&232.7333590&-81.5463100&0.0035&0.0045&54285.61403&0.578059536&0.098753214\\
1&11&F9YM0108M&15.8&280.612&-11.4354866&-23.3057775&0.0037&0.0040&54285.61709&0.579074299&0.098937542\\
1&5&F9YM0109M&14.44&280.612&232.7329783&-81.5462846&0.0036&0.0053&54285.61858&0.577974028&0.098734923\\
1&1&F9YM010AM&8.2&280.612&-5.3805110&2.5179985&0.0028&0.0059&54285.61983&0.579044335&0.098811176\\
1&5&F9YM010BM&14.42&280.612&232.7339517&-81.5449234&0.0033&0.0038&54285.62132&0.577923291&0.098723367\\
1&5&F9YM010CM&14.43&280.612&232.7326276&-81.5462801&0.0037&0.0041&54285.62263&0.577899611&0.098717663\\
1&9&F9YM010DM&13.97&280.612&163.6020574&-32.1466724&0.0033&0.0034&54285.62382&0.578243003&0.098638334\\
1&11&F9YM010EM&15.79&280.612&-11.4345429&-23.3038134&0.0036&0.0038&54285.62524&0.578926197&0.098902322\\
1&1&F9YM010FM&8.2&280.612&-5.3807179&2.5170693&0.0024&0.0035&54285.62623&0.578930526&0.098782607\\
1&5&F9YM010GM&14.42&280.612&232.7327101&-81.5468546&0.0036&0.0031&54285.62767&0.577812914&0.098694251\\
2&1&F9YM0301M&8.2&253.022&1.8980570&-45.9885308&0.0015&0.0015&54297.38964&0.400934043&0.038705436\\
2&5&F9YM0302M&14.42&253.022&251.8340113&-10.0603051&0.0026&0.0017&54297.39127&0.399679689&0.038708685\\
...&...&...&...&...&...&...&...&...&...&...&...
\enddata
\tablenotetext{a}{Set (orbit) number, star number (\#1 = HD 202206; reference star numbers same as Table~\ref{tbl-1}), \HSTs orbit and target identifier, V magnitude from FGS measure, spacecraft +V3 axis roll angle as defined in Chapter 2, FGS Instrument 
Handbook \citep{Nel15a}, OFAD X and Y positions in arcsec, position measurement errors in arcsec, time of observation = JD - 2400000.5, RA and DEC parallax factors. We provide a complete table in the
electronic version of this paper.}
\end{deluxetable}
\end{landscape}

\begin{deluxetable}{rlllll}
\tablewidth{0in}
\tablecaption{ Astrometric Reference Stars\label{tbl-1}}
\tablehead{
\colhead{ID}
&\colhead{RA\tablenotemark{a}~~~ (J2000.0)~}&
\colhead{DEC\tablenotemark{a}}&
\colhead{V\tablenotemark{b}}
}
\startdata
5&318.666441&-20.778502&14.30\\
6&318.705720&-20.830115&14.54\\
9&318.689335&-20.788494&13.95\\
10&318.756629&-20.809268&15.98\\
11& 318.740855\tablenotemark{c}& -20.782022\tablenotemark{c}&15.92\\
\enddata
\tablenotetext{a}{Positions from PPMXL \citep{Roe10}, J2000.}
\tablenotetext{b}{V magnitude, this paper.}
\tablenotetext{c}{Position from GSC2.3 \citep{Las08}.}
\end{deluxetable}

\begin{deluxetable}{ccccccc}
\tablewidth{0in}
\tablecaption{Visible and Near-IR Photometry \label{tbl-IR}}
\tablehead{\colhead{ID}&
\colhead{$V$} &
\colhead{$B-V$} &
\colhead{$K$} &
\colhead{$(J-H)$} &
\colhead{$(J-K)$} &
\colhead{$(V-K)$} 
}
\startdata
1&8.08$\pm$0.03&0.72$\pm$0.03&6.485$\pm$0.023&0.283$\pm$0.031&0.365$\pm$0.035&1.60$\pm$0.04\\
5&14.30 0.03&0.73 0.10\tablenotemark{a}&12.442 0.023&0.423 0.034&0.484 0.032&1.86 0.04\\
6&14.00  0.03&0.74 0.05&12.465 0.026&0.456 0.033&0.537 0.035&2.08 0.04\\
9&13.95 0.03&0.70 0.05&12.318 0.025&0.330 0.034&0.385 0.036&1.63 0.04\\
10&15.98 0.03&0.65 0.09&14.150 0.066&0.255 0.064&0.497 0.073&1.83 0.07\\
11&15.92 0.10 &0.87 0.09&&&&\\
\hline
\enddata
\tablenotetext{a}{Estimated from 2MASS photometry.}
\end{deluxetable}

\begin{deluxetable}{ccccccc}
\tablewidth{0in}
\tablecaption{Astrometric Reference Star Initial
Spectrophotometric Parallaxes \label{tbl-SPP}}
\tablehead{\colhead{ID}& \colhead{Sp. T.\tablenotemark{a}}&
\colhead{V} & \colhead{M$_V$} &\colhead{m-M}& \colhead{A$_V$}&
\colhead{$\pi_{abs}$(mas)}} 
\startdata
5&K1.5V&14.30&6.3&8.0&0.00&2.5$\pm$0.6\\
6&K0V&14.54&5.9&8.6&0.00&1.9 0.4\\
9&ÊÊÊG5V ÊÊ&13.95&5.1&8.9&0.06&1.6 0.4\\
10&ÊÊÊG5V ÊÊ&15.98&5.1&10.9&0.00&0.7 0.2\\
\enddata 
\tablenotetext{a}{Spectral types and luminosity class estimated from classification spectra, colors, and 
reduced 
proper motion diagram (Figures~\ref{fig-CCD} and~\ref{fig-RPM}).}\\
\end{deluxetable}

\begin{deluxetable}{ccrr}
\tablewidth{0in}
\tablecaption{HD 202206 and Reference Star Relative Positions\tablenotemark{a}   \label{tbl-POS}}
\tablecolumns{6} 
\centering
\tablehead{\colhead{Star}&\colhead{V}& \colhead{$\xi$ } &  \colhead{$\eta$}}
\startdata
1&8.08&0.19070$\pm$0.00013&20.77968$\pm$0.00014\\
5&14.3&-249.19116 0.00011&60.23407 0.00007\\
6&14.54&-117.09436 0.00046&-125.45195 0.00032\\
9\tablenotemark{b}&13.95&-172.23627 0.00018&24.24109 0.00015\\
10&15.98&54.06826 0.00052&-50.63166 0.00047\\
11&15.92&1.50403 0.00018&47.38927 0.00017\\
\enddata
\tablenotetext{a}{~Units are arc seconds, rolled to RA ($\xi$) and DEC ($\eta$), epoch 2008.4085 (J2000). Roll uncertainty $\pm0\fdg06$.}
\tablenotetext{b}{RA = 318.689335, DEC = -20.788494, J2000}

\end{deluxetable}

\begin{deluxetable}{ccrrrr}
\tablewidth{0in}
\tablecaption{Reference Star Final Proper Motions, Parallaxes, and Absolute Magnitudes \label{tbl-PM}}
\tablehead{
\colhead{ID}&
\colhead{V} &
\colhead{$\mu_\alpha$\tablenotemark{a}} &
\colhead{$\mu_\delta$\tablenotemark{a}}&
\colhead{$\pi_{abs}$} &
\colhead{M$_V$}
}
\startdata 
5&14.3&-6.66$\pm$0.10&-22.63$\pm$0.11&2.35$\pm$0.13&6.15$\pm$0.05\\
6&14.54&3.00 0.45&-9.43 0.50&1.93 0.10&5.96 0.05\\
9&13.95&-11.29 0.16&-14.28 0.17&1.74 0.13&5.09 0.07\\
10&15.98&-13.69 0.40&-8.05 0.42&0.67 0.04&5.10 0.05\\
11&15.92&3.95 0.17&-0.72 0.19&1.08 0.06&6.08 0.05\\
\enddata
\tablenotetext{a}{Proper motions are relative  in mas
yr$^{-1}$. Parallax in mas. }
\end{deluxetable}


\begin{deluxetable}{ll}
\tablecaption{Reference Frame Statistics, HD 202206 Parallax, and Proper Motion\label{tbl-SUM}}
\tablewidth{0in}
\tablehead{\colhead{Parameter} &  \colhead{Value} }
\startdata
Study duration  &2.91 y  \\
number of observation sets    &   31  \\
reference star $\langle V\rangle$ &  14.94     \\
reference star $\langle (B-V) \rangle$ &0.79   \\
{\it HST}~Absolute $\pi$& 21.96 $\pm$ 0.12    mas \\
~~~~~~~Relative  $\mu_\alpha$& -41.54 $\pm$ 0.11 mas yr$^{-1}$\\
~~~~~~~Relative  $\mu_\delta$&  -117.87 $\pm$ 0.11  mas yr$^{-1}$\\
~~~~~~~$\vec{\mu} = 124.98$ mas  yr$^{-1}$\\
~~~~~~~P.A. = $199\fdg4$\\
\G~DR1 Absolute $\pi$& 21.94 $\pm$ 0.26     mas \\
~~~~~~~Absolute  $\mu_\alpha$& -39.22 $\pm$ 0.07 mas yr$^{-1}$\\
~~~~~~~Absolute  $\mu_\delta$&  -120.29 $\pm$ 0.04  mas yr$^{-1}$\\
~~~~~~~$\vec{\mu} = 126.53$ mas  yr$^{-1}$\\
~~~~~~~P.A. = $198\fdg1$\\
{\it HIP}07 Absolute $\pi$& 22.06 $\pm$ 0.82     mas \\
~~~~~~~Absolute  $\mu_\alpha$& -38.40 $\pm$ 0.94 mas yr$^{-1}$\\
~~~~~~~Absolute  $\mu_\delta$&  -119.81 $\pm$ 0.37  mas yr$^{-1}$\\
~~~~~~~$\vec{\mu} = 125.81$ mas  yr$^{-1}$\\
~~~~~~~P.A. = $197\fdg8$\\

\enddata
\end{deluxetable}

\begin{deluxetable}{ccrlrl}
\tablecaption{Orbital Elements for the \HDs B and c Perturbations\label{tbl-ORB}}
\tablewidth{6in}
\tablehead{
\colhead{Parameter} &  
\colhead{Units} &
\colhead{B} &
\colhead{err} &
\colhead {c} &
\colhead{err}
		}
\startdata
P&days&256.33&0.02&1260&11\\
P&years&0.70180&0.00005&3.45&0.03\\
$T_0$&JD-2400000&52176.14&0.12&53103&452\\
e& -&0.432&0.001&0.22&0.03\\
K&\kms&0.567&0.001&0.041&0.001\\
i&$\arcdeg$&10.9&0.8&7.7&1.1\\
$\omega$&$\arcdeg$&161.9&0.2&280&4\\
$\Omega$&$\arcdeg$&121&4&91&11\\
$\alpha$&mas&1.40&0.10&0.76&0.11\\
\hline
Derived Parameters &&&&&\\
$\alpha$&AU&0.064&0.005&0.035&0.005\\
a&AU&0.83&&2.41&\\
a&mas&18.2&&52.9&\\
\msini&\mjupe&17.7&&2.3&\\
\m&\mjupe&93.6&$^{+7.7}_{-6.6}$&17.9&$^{+2.9}_{-1.8}$\\
\m&\msun&0.089&&0.017&\\
\hline
Stability Parameters &&&&&\\
$P_c/P_B$&-&4.92& 0.04&&\\
$\Phi$\tablenotemark{a}&$\arcdeg$&6& 2&&\\
\enddata
\tablenotetext{a}{Mutual inclination from Equation~\ref{MI}}
\end{deluxetable}

\begin{deluxetable}{ccrlrl}
\tablecaption{Orbital Elements with 5:1 MMR Prior\label{tbl-resRES}}
\tablewidth{6in}
\tablehead{
\colhead{Parameter} &  
\colhead{Units} &
\colhead{B} &
\colhead{err} &
\colhead {c} &
\colhead{err}
		}
\startdata
P&days&256.31&0.02&1278&6\\
P&years&0.70174&0.00004&3.50&0.02\\
$T_0$&JD-2400000&52176.10&0.11&53109&223\\
e& -&0.432&0.001&0.20&0.03\\
K&\kms&0.567&0.001&0.041&0.001\\
i&$\arcdeg$&10.8&0.8&7.7&1.1\\
$\omega$&$\arcdeg$&161.9&0.2&280&4\\
$\Omega$&$\arcdeg$&121&4&100&9\\
$\alpha$&mas&1.40&0.10&0.76&0.11\\
\hline
Derived Parameters &&&&&\\
$\alpha$&AU&0.064&0.005&0.035&0.005\\
a&AU&0.83&&2.43&\\
a&mas&18.2&&53.4&\\
\msini&\mjupe&17.7&&2.3&\\
\m&\mjupe&93.9&$^{+7.6}_{-6.5}$&18.0&$^{+2.9}_{-2.0}$\\
\m&\msun&0.090&&0.017&\\
\hline
Stability Parameters &&&&&\\
Period ratio, $P_c/P_B$&-&4.99& 0.02&&\\
$Ppdiff$&days & 3 &5 &&\\
$\Phi$\tablenotemark{a}&$\arcdeg$&4& 2&&\\
\enddata
\tablenotetext{a}{Mutual inclination from Equation~\ref{MI}}
\end{deluxetable}

%
%

\clearpage

\clearpage

\clearpage
\clearpage

\clearpage

\end{document}